\documentclass[11pt]{article}
\usepackage[a4paper,margin=2.5cm]{geometry}
\usepackage{graphicx}
\usepackage{authblk}
\newcommand{\articletype}[1]{}
\usepackage[
    colorlinks=true,
    linkcolor=blue,
    citecolor=blue,
    urlcolor=blue
]{hyperref}
\usepackage{amsmath}
\usepackage{amssymb}
\usepackage{dcolumn}
\usepackage{bm}
\newcommand{\sigmamax}{\ensuremath{\sigma_{\mathrm{max}}}}
\newcommand{\sigmass}{\ensuremath{\sigma_{\mathrm{ss}}}}
\newcommand{\gammadot}{\ensuremath{\dot{\gamma}}}
\newcommand{\Ts}{\ensuremath{T_{\mathrm{s}}}}
\usepackage{xcolor}

\newcommand{\Tp}{\ensuremath{T_{\mathrm{p}}}}
\newcommand{\sigmay}{\ensuremath{\sigma_{\mathrm{y}}}}
\newcommand{\Kss}{\ensuremath{K_{\mathrm{ss}}}}
\newcommand{\nss}{\ensuremath{n_{\mathrm{ss}}}}
\newcommand{\Ass}{\ensuremath{A_{\mathrm{ss}}}}
\newcommand{\Kmax}{\ensuremath{K_{\mathrm{max}}}}
\newcommand{\nmax}{\ensuremath{n_{\mathrm{max}}}}
\newcommand{\Amax}{\ensuremath{A_{\mathrm{max}}}}
\newcommand{\ath}{\mathrm{ath}}
\newcommand{\AQS}{\mathrm{AQS}}

\begin{document}

\articletype{Article}

\title{Thermal softening of the yield stress in glasses with different stability}

\author{Tianze Wan and Ludovic Berthier}

\affil{Gulliver, CNRS UMR 7083, ESPCI Paris, PSL Research University, 75005 Paris, France}

\maketitle

\begin{abstract}
The yielding of amorphous solids has been extensively studied in the athermal limit, but the influence of thermal fluctuations is much less understood, in particular for glasses of different stability. Using molecular dynamics simulations of a model glass where the swap Monte Carlo algorithm can be employed, we study the yield stress at finite temperature and shear rate, both in steady flow and at shear start-up, over a broad range of glass stability. Thermal fluctuations soften both the dynamic yield stress in steady state and the start-up yield stress. In all cases, the temperature dependence is compatible with the $T^{2/3}$ law expected for thermally activated yielding near a saddle-node instability, with no systematic dependence of the exponent on stability or shear rate. Extrapolations to zero temperature recover the athermal yield stresses. The amplitude of the softening at start-up, by contrast, depends strongly on stability: highly stable glasses are much less sensitive to temperature than poorly annealed ones. A description of yielding at finite temperature must therefore account explicitly for the stability of the glass.
\end{abstract}

\section{Introduction}

\label{sec:Intro}

Amorphous solids, from metallic and colloidal glasses to foams, span a broad range of length and time scales~\cite{berthierTheoretical2011,nicolasDeformation2018}. Despite this diversity, their mechanical response shares universal features: they respond elastically to small deformations, and yield and flow plastically at large deformations~\cite{bonnYield2017,berthierYielding2025}. Understanding these features is important both for a unified theoretical description of amorphous solids and for applications.

The way an amorphous solid yields depends strongly on how it was prepared. Well-annealed glasses yield in a brittle manner, with a sharp stress drop and the formation of a shear band, whereas poorly annealed glasses yield in a ductile manner, with a smooth evolution of the stress~\cite{ozawaRandom2018,yehGlass2020,lampBrittle2022,bhaumikAvalanches2022,bhaumikYielding2022}. In the athermal quasi-static (AQS) limit, the two regimes are separated by a random critical point~\cite{ozawaRandom2018}, and a finite shear rate smooths the transition~\cite{singhBrittle2020}. Less is known about the role of temperature, which is unavoidable in experiments.

To separate the role of thermal fluctuations during deformation from that of the glass structure, two temperatures need to be distinguished~\cite{falkThermal2004,falkDeformation2011}: the temperature $\Ts$ at which the glass is deformed, which controls thermally activated rearrangements and plastic events, and the parent temperature $\Tp$ from which the glass was prepared, which sets its stability. The dependence of yielding on $\Ts$ is usually interpreted in terms of thermal activation. Johnson and Samwer found an approximately universal $\Ts^{2/3}$ reduction of the yield stress across a broad range of metallic glasses, which they attributed to the thermal activation of shear transformation zones before the barrier created by the loading vanishes~\cite{johnsonUniversal2005}. Chattoraj \textit{et al.}~derived a corresponding additive thermal correction to the steady-state flow stress~\cite{chattorajUniversal2010}, and the same picture was used to describe yielding at start-up as the thermally activated nucleation of a shear band~\cite{dasguptaMicroscopic2012,dasguptaDerivation2013}. Cheng and Ma distinguished the intrinsic strength associated with homogeneous nucleation from the lower strength associated with heterogeneous nucleation at ``stress concentrators'', with distinct temperature dependencies~\cite{chengIntrinsic2011}. Other activation models predict different exponents, such as the $\Ts^{1/2}$ dependence of Argon-type models~\cite{schuhMechanical2007}, and deviations from the Johnson--Samwer law have been reported~\cite{songUniversal2022,dasguptaDerivation2013,dubachConstitutive2009}. More recently, thermal activation has also been incorporated into elastoplastic descriptions of yielding, where it competes with mechanical driving and rounds the athermal yielding transition~\cite{popovicThermally2021,ferreroYielding2021,korchinskiThermally2024}.

Numerical tests of these predictions exist for steady flow~\cite{chattorajUniversal2010} and for start-up~\cite{chengIntrinsic2011,rottlerShear2003}. For start-up, however, they rely on conventionally prepared glasses, whose stability can only be varied over a narrow range. Yet, unlike in steady flow, the glass at start-up is not in a stationary state. The energy barriers that thermal fluctuations help to overcome are those of the initial glass, and they depend on how it was prepared. Whether glass stability changes the exponent of the thermal softening, its amplitude, or both, has therefore, to our knowledge, not been tested systematically. The swap Monte Carlo algorithm now makes it possible to equilibrate numerical glass-formers far below the conventional glass-transition regime~\cite{ninarelloModels2017,berthierEfficient2019}, and thus to prepare glasses over a very broad range of stability, as exploited in the athermal limit in Refs.~\cite{ozawaRandom2018,yehGlass2020,singhBrittle2020}.

Here, we use the swap Monte Carlo algorithm to study yielding at finite temperature and finite shear rate over a broad range of glass stability, and we analyze both steady flow and shear start-up in the same model. We find that thermal fluctuations soften both the dynamic yield stress and the start-up yield stress, with a temperature dependence compatible with the $\Ts^{2/3}$ law in all cases, and no systematic dependence of the exponent on stability or shear rate. The amplitude of the softening at start-up, by contrast, depends strongly on stability. We find that stable glasses are much less sensitive to temperature than poorly annealed ones.

The article is organized as follows. Section~\ref{sec:Model} presents the model and the simulation protocols. Section~\ref{sec:SteadyState} analyzes the steady flow and the temperature dependence of the dynamic yield stress. Section~\ref{sec:Startup} turns to shear start-up: after describing the stress--strain curves, we study the shear-rate and temperature dependence of the start-up yield stress, and compare its thermal softening with that of the steady flow. Section~\ref{sec:Conclusions} summarizes our results and discusses some perspectives.

\section{Model and methods}

\label{sec:Model}

We simulate a three-dimensional size-polydisperse system of $N=1500$ particles at number density $\rho=1$ in a cubic box with periodic boundary conditions~\cite{ninarelloModels2017,berthierEfficient2019}. Particles $i$ and $j$ separated by a distance $r_{ij}$ interact via the soft repulsive potential
\begin{equation}
    \label{SoftPotential}
    \begin{aligned}
        u(r_{ij})/\epsilon &= \left(\frac{d_{ij}}{r_{ij}}\right)^{12} + c_0 + c_2\left(\frac{r_{ij}}{d_{ij}}\right)^{2} + c_4\left(\frac{r_{ij}}{d_{ij}}\right)^{4},\\
        d_{ij} &= \frac{d_i+d_j}{2}\left( 1-0.2\,|d_i-d_j| \right),
    \end{aligned}
\end{equation}
where $d_i$ is the diameter of particle $i$ and $\epsilon$ sets the energy scale. The non-additive mixing rule for $d_{ij}$ prevents crystallization of the polydisperse mixture~\cite{ninarelloModels2017}. The constants $c_0=-28\,r_{\mathrm{cut}}^{-12}$, $c_2=48\,r_{\mathrm{cut}}^{-14}$ and $c_4=-21\,r_{\mathrm{cut}}^{-16}$ ensure that the potential and its first two derivatives vanish at the cutoff distance $r_{ij}=r_{\mathrm{cut}}\,d_{ij}$, with $r_{\mathrm{cut}}=1.25$. The diameters are drawn from the distribution $P(d)=C/d^{3}$ for $0.73\leq d\leq 1.62$, where $C$ is a normalization constant. All particles have the same mass $m$. Lengths, times and temperatures are expressed in units of $\bar{d}=\int d\,P(d)\,\mathrm{d}d$, $\sqrt{m\bar{d}^{2}/\epsilon}$ and $\epsilon/k_{\mathrm{B}}$, respectively.

We integrate the \textsc{SLLOD} equations of motion with \textsc{LAMMPS}~\cite{LAMMPS}, using a time step $\Delta t=0.01$. A constant shear rate $\gammadot$ is imposed by deforming the simulation box and remapping particle positions and velocities accordingly, which is equivalent to Lees--Edwards boundary conditions. The strain is $\gamma=\gammadot\, t$. The shear stress is the $xy$ component of the stress tensor,
$\sigma = \frac{1}{V}\sum_{i=1}^{N} m\, v_{ix}v_{iy}+\frac{1}{V}\sum_{i<j} r_{ij,x}f_{ij,y}$, where $V$ is the volume, $\mathbf{v}_i$ is the velocity of particle $i$, and the second sum runs over interacting pairs, with $\mathbf{r}_{ij}$ their separation and $\mathbf{f}_{ij}$ the corresponding force. The two terms are the kinetic and virial contributions to the stress. Thermostatted \textsc{SLLOD} dynamics has been reported to suppress strain localization~\cite{lampBrittle2022,shrivastavHeterogeneous2016}. However, we do observe shear bands in our simulations of stable glasses, similar to what was reported in simulations at finite temperature~\cite{shiEvaluation2007} and in athermal conditions~\cite{parmarStrain2019}.

The temperature $\Ts$ is imposed by a Nos\'e--Hoover thermostat acting on the kinetic temperature. Two technical points require care. First, the thermostat has its own relaxation time. It can make the kinetic temperature oscillate around its target value, and it interferes with the driven dynamics if it approaches the shear time scale $1/\gammadot$, which here ranges from $2.77\times10^{2}$ to $1.93\times10^{6}$. With the commonly used damping time of 100 time steps, we observed oscillations of the kinetic energy. We therefore use a damping time of 5 time steps, i.e.\ $0.05$, which suppresses them and remains far below $1/\gammadot$. Second, the thermostat must remove the heat produced by the shear flow, which sets an upper bound on $\gammadot$. Above it, the kinetic temperature drifts away from $\Ts$~\cite{ruiz-francoEffect2018}. Varnik \textit{et al.}~estimated this bound by comparing the heating rate with the time needed for heat to cross the system~\cite{varnikStudy2004}. Here, we determine it empirically for each $\Ts$, and only use shear rates for which the kinetic energy per particle, $E_{\mathrm{k}}$, keeps its equilibrium statistics. The relative deviations of $\langle E_{\mathrm{k}}\rangle$ from $3\Ts/2$ and of $\mathrm{Var}(E_{\mathrm{k}})$ from $3\Ts^{2}/(2N)$ remain below $5\%$. The same criterion sets the lowest temperatures we study: at very small $\Ts$, the kinetic energy is no longer well controlled by the thermostat.

Initial configurations are equilibrated at the parent temperature $\Tp$ using the swap Monte Carlo algorithm, with the parameters of Ref.~\cite{berthierEfficient2019}. Swap Monte Carlo gives access to equilibrium configurations over a broad range of $\Tp$, and therefore to glasses of widely different stability. We use $\Tp = 0.062$, $0.075$, $0.085$, $0.092$, $0.1$ and $0.125$. For comparison, the mode-coupling crossover temperature of this model is $T_{\mathrm{MCT}}\approx0.095$ and its estimated experimental glass transition temperature is $T_{\mathrm{g}}\approx0.056$~\cite{guiselinMicroscopic2022}. The lowest $\Tp$ thus produces very stable glasses, and the highest $\Tp$ poorly annealed ones. For the highest parent temperature, $\Tp=0.125$, the stress overshoot defined below is weak and its maximum becomes less well defined. This value therefore marks the upper limit of the range of $\Tp$ we can analyze quantitatively, and we only include it in the comparison of Sec.~\ref{sec:Comp}, not in the quantitative analysis of Secs.~\ref{sec:Rate} and~\ref{sec:Softening}.

\begin{figure}[t]
    \centering
    \includegraphics[width=0.8\textwidth]{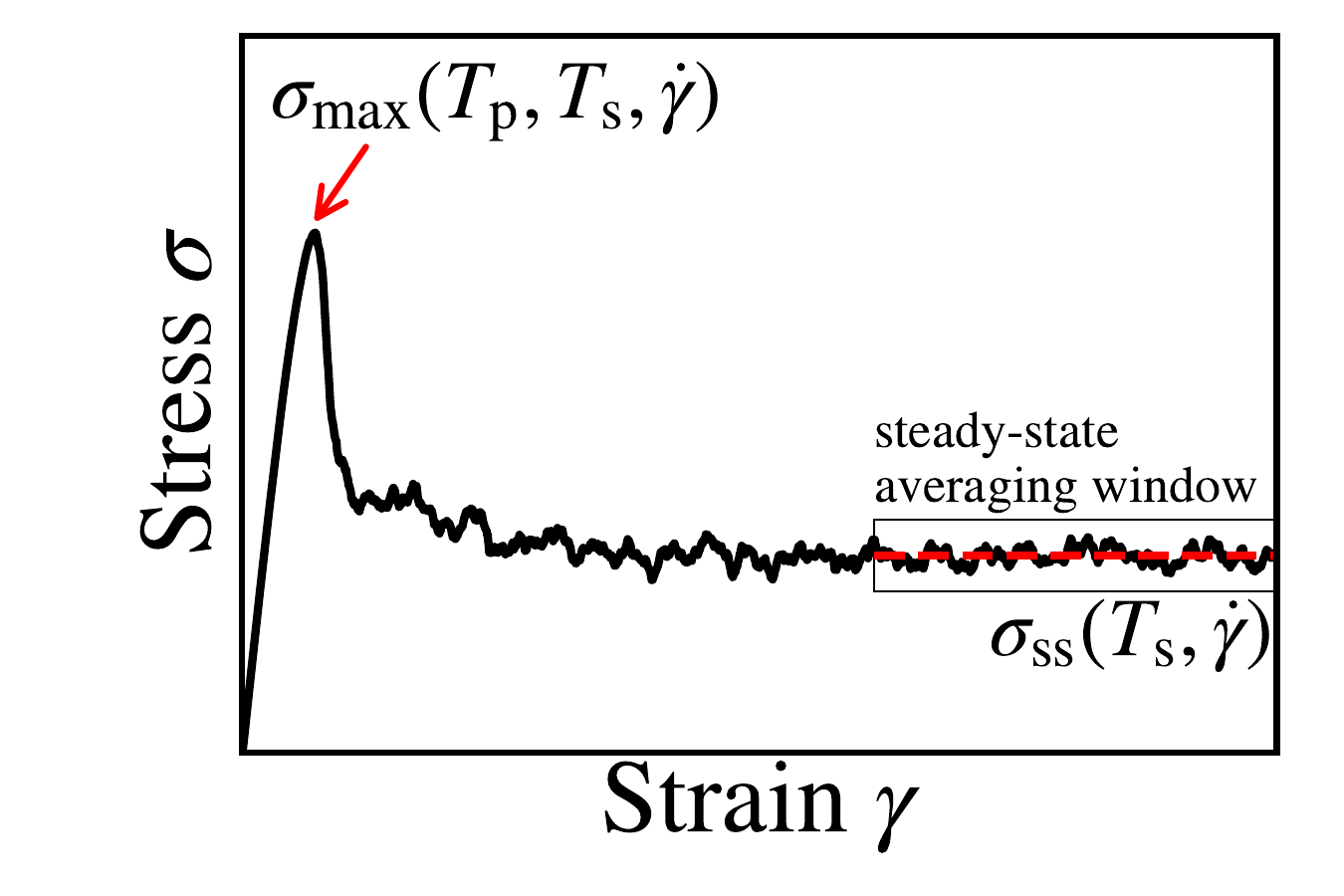}
    \caption{Schematic stress--strain curve during shear start-up. The stress overshoot is quantified by the start-up yield stress $\sigmamax(\Ts,\Tp,\gammadot)$, the maximum of the curve. At large strain, the stress fluctuates around the steady-state stress $\sigmass(\Ts,\gammadot)$ (red dashed line), which does not depend on the parent temperature $\Tp$. The framed region indicates the strain window used to measure $\sigmass$.}
    \label{flow_sketch}
\end{figure}

For shear start-up, each equilibrium configuration is first quenched to its inherent structure using the FIRE algorithm~\cite{bitzekStructural2006}. Velocities drawn from the Maxwell--Boltzmann distribution at temperature $\Ts$ are then assigned, and the system evolves for $10^4$ time units, long enough to equilibrate the velocities but too short to modify the structure inherited from $\Tp$. During this waiting time, the mean-squared displacement remains on its plateau, and the time-averaged potential energy $\langle U\rangle$ exceeds the inherent-structure energy $E_{\mathrm{IS}}$ by $\frac{3}{2}(N-1)\Ts$, as expected for vibrations within the initial energy basin. Therefore, the velocities equilibrate but the structure does not relax. Shear is then applied at temperature $\Ts$ and shear rate $\gammadot$. On the timescale associated with the lowest shear rates, thermal aging can occur even in the absence of shear. However, the corresponding change in $E_{\mathrm{IS}}$ remains small compared with the differences in $E_{\mathrm{IS}}$ between parent temperatures. Figure~\ref{flow_sketch} sketches a typical stress--strain curve. The stress first increases with strain, passes through a maximum, and then relaxes towards a steady state at large strain. We call this non-monotonic evolution the stress overshoot, and quantify it by the start-up yield stress $\sigmamax = \sigmamax(\Ts,\Tp,\gammadot)$, defined as the maximum of the stress--strain curve averaged over 20 to 40 independent initial configurations.

At large strain, the stress fluctuates around the steady-state stress $\sigmass =  \sigmass(\Ts,\gammadot)$. In steady state the system has lost memory of its initial configuration, so $\sigmass$ does not depend on $\Tp$. We therefore measure it by shearing equilibrium configurations prepared at $\Tp=0.085$, without the quench to inherent structures. For each state point, $\sigmass$ is obtained from a single configuration by averaging the stress over the window $2\leq\gamma\leq20$, whose duration provides sufficient statistics. Overall, we study temperatures $\Ts$ between $4.45\times10^{-5}$ and $8.5\times10^{-2}$ and shear rates $\gammadot$ between $10^{-5}$ and $3.61\times10^{-3}$ in steady flow, and $\Ts$ between $4.39\times10^{-4}$ and $4.31\times10^{-2}$ and $\gammadot$ between $5.18\times10^{-7}$ and $1.64\times10^{-4}$ at start-up. Uncertainties on all fitted parameters are estimated from 2000 bootstrap resamples of the data entering each fit. The extension of each error bar is the width of the central 95\% interval of the bootstrap distribution.

\section{Steady flow and the dynamic yield stress}

\label{sec:SteadyState}

We first characterize the steady-state flow. Since it does not depend on the preparation of the glass, it provides a reference for the start-up response studied in Sec.~\ref{sec:Startup}.

\begin{figure}[t]
    \centering
    \includegraphics[width=1\textwidth]{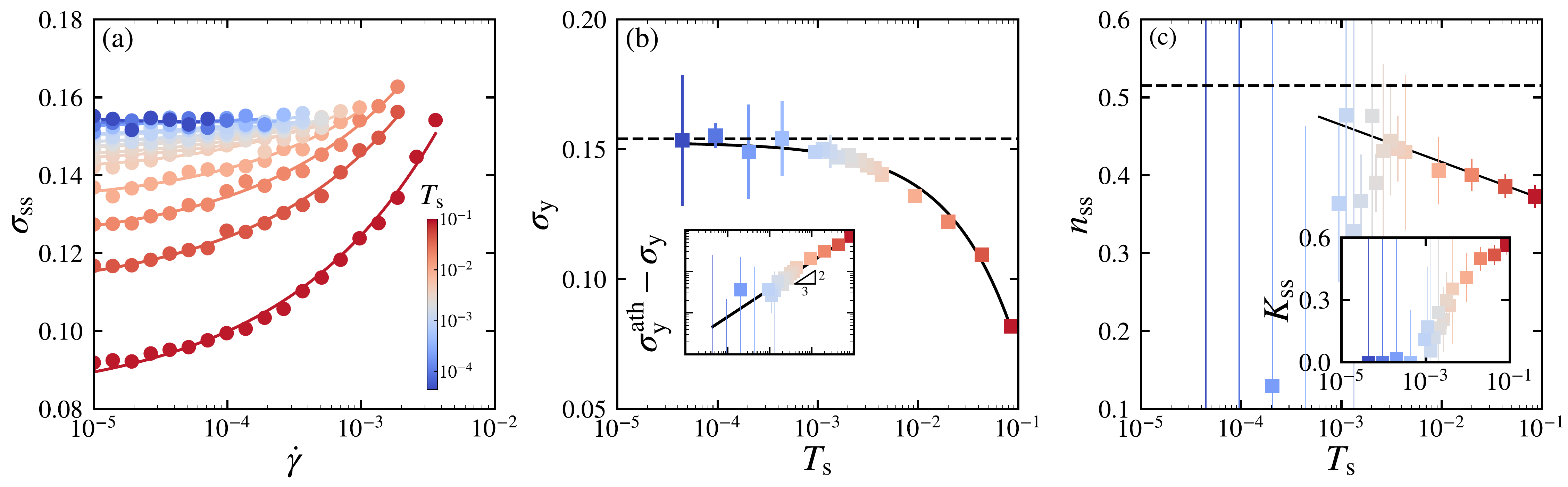}
    \caption{Steady-state rheology at finite temperature.
            (a) Flow curves $\sigmass(\gammadot)$ for different temperatures $\Ts$. Solid lines are fits to the Herschel--Bulkley form, Eq.~\eqref{HB}.
            (b) Dynamic yield stress $\sigmay$ obtained from these fits, as a function of $\Ts$. The dashed line indicates the AQS value $\sigmay^{\AQS}=0.154$ for the same model~\cite{singhBrittle2020}. The inset shows $\sigmay^{\ath}-\sigmay$ as a function of $\Ts$. The black solid lines in the main panel and in the inset are fits to Eq.~\eqref{lemaitre_fit}.
            (c) Herschel--Bulkley exponent $\nss$ as a function of $\Ts$. The inset shows the prefactor $\Kss$. The dashed line indicates the AQS value $\nss^{\AQS}=0.515$ for the same model~\cite{singhBrittle2020}, and the black solid line is a guide to the eye.
            Error bars are bootstrap estimates (Sec.~\ref{sec:Model}) and become very large at very low $\Ts$ where the yield stress varies very little, see (a).}
    \label{flow_curve}
\end{figure}

Figure~\ref{flow_curve}(a) shows the flow curves $\sigmass(\gammadot)$ at different temperatures $\Ts$. At low shear rates, $\sigmass$ approaches a finite plateau, and it increases at larger $\gammadot$. The flow curves are well described by the Herschel--Bulkley form~\cite{herschelKonsistenzmessungen1926},
\begin{equation}
    \label{HB}
    \sigmass(\gammadot)=\sigmay+\Kss\gammadot^{\nss},
\end{equation}
shown by the solid lines in Fig.~\ref{flow_curve}(a). Here, $\sigmay$ is the dynamic yield stress~\cite{bonnYield2017}, $\Kss$ is a prefactor, and $\nss$ is the Herschel--Bulkley exponent, typically found in the range $0.4$--$0.6$ for soft glassy materials~\cite{bonnYield2017,singhBrittle2020}. Since $\sigmay$ is obtained by extrapolating the fits to $\gammadot\to 0$, it describes the zero-rate limit of the flow curves over the accessible range of shear rates. Therefore $\sigmay$ only depends on the temperature $\Ts$. 

Figure~\ref{flow_curve}(b) shows that $\sigmay$ decreases with increasing $\Ts$. Thermal fluctuations soften the steady flow. At low $\Ts$, $\sigmay$ approaches the value $\sigmay^{\AQS} \simeq 0.154$ obtained for the same model in the athermal quasi-static (AQS) limit~\cite{singhBrittle2020}. Chattoraj \textit{et al.}~showed that the thermally activated triggering of local plastic instabilities near a saddle-node bifurcation lowers the steady-state stress by an amount proportional to $\Ts^{2/3}$, up to logarithmic corrections in $\gammadot$~\cite{chattorajUniversal2010}. Since this temperature dependence is established for steady flow, we fix the exponent to $2/3$ and, neglecting logarithmic corrections, describe our data by
\begin{equation}
\label{lemaitre_fit}
    \sigmay(\Ts)=\sigmay^{\ath}-\Ass\Ts^{2/3},
\end{equation}
where $\sigmay^{\ath}$ is the athermal limit of $\sigmay$ and $\Ass$ is the thermal-softening amplitude. Whether the same exponent also holds at start-up, where the stability of the glass matters, is the question addressed in Sec.~\ref{sec:Startup}. Equation~\eqref{lemaitre_fit} describes the data well, as shown by the solid line in Fig.~\ref{flow_curve}(b), with $\Ass=0.37\pm0.07$. The fitted value $\sigmay^{\ath}=0.153\pm0.002$ agrees with $\sigmay^{\AQS}$, and fixing $\sigmay^{\ath}=\sigmay^{\AQS}$ gives the same result. The inset of Fig.~\ref{flow_curve}(b), which shows $\sigmay^{\ath}-\sigmay$ as a function of $\Ts$, confirms the $\Ts^{2/3}$ dependence, consistent with thermal activation near a saddle-node instability.

The other two Herschel--Bulkley parameters are shown in Fig.~\ref{flow_curve}(c). At the higher temperatures, where they are well determined, $\nss$ decreases with increasing $\Ts$, consistent with experimental observations~\cite{caggioniVariations2020}, and its trend points towards the athermal value $\nss^{\AQS} \simeq 0.515$ obtained for the same model~\cite{singhBrittle2020}. The value of the Herschel--Bulkley exponent has been related both to avalanche dynamics and to the shape of the local yielding potential~\cite{linMicroscopic2018,ferreroElastic2019}. The prefactor $\Kss$ (inset) decreases more strongly with $\Ts$. Together, these variations reduce the rate-dependent part of the stress at higher temperature. At very low temperatures, $\Ts<10^{-3}$, by contrast, the flow curves are nearly flat over the accessible range of shear rates. The term $\Kss\gammadot^{\nss}$ is then small, and $\nss$ and $\Kss$ are poorly constrained, as reflected by the error bars in Fig.~\ref{flow_curve}(c) that become very large at very low $\Ts$. 

\section{Shear start-up}

\label{sec:Startup}

We turn to shear start-up, where the response now depends explicitly on the initial stability of the glass.

\subsection{Stress--strain curves}

\label{sec:Curves}

\begin{figure}[htbp]
    \centering
    \includegraphics[width=1\textwidth]{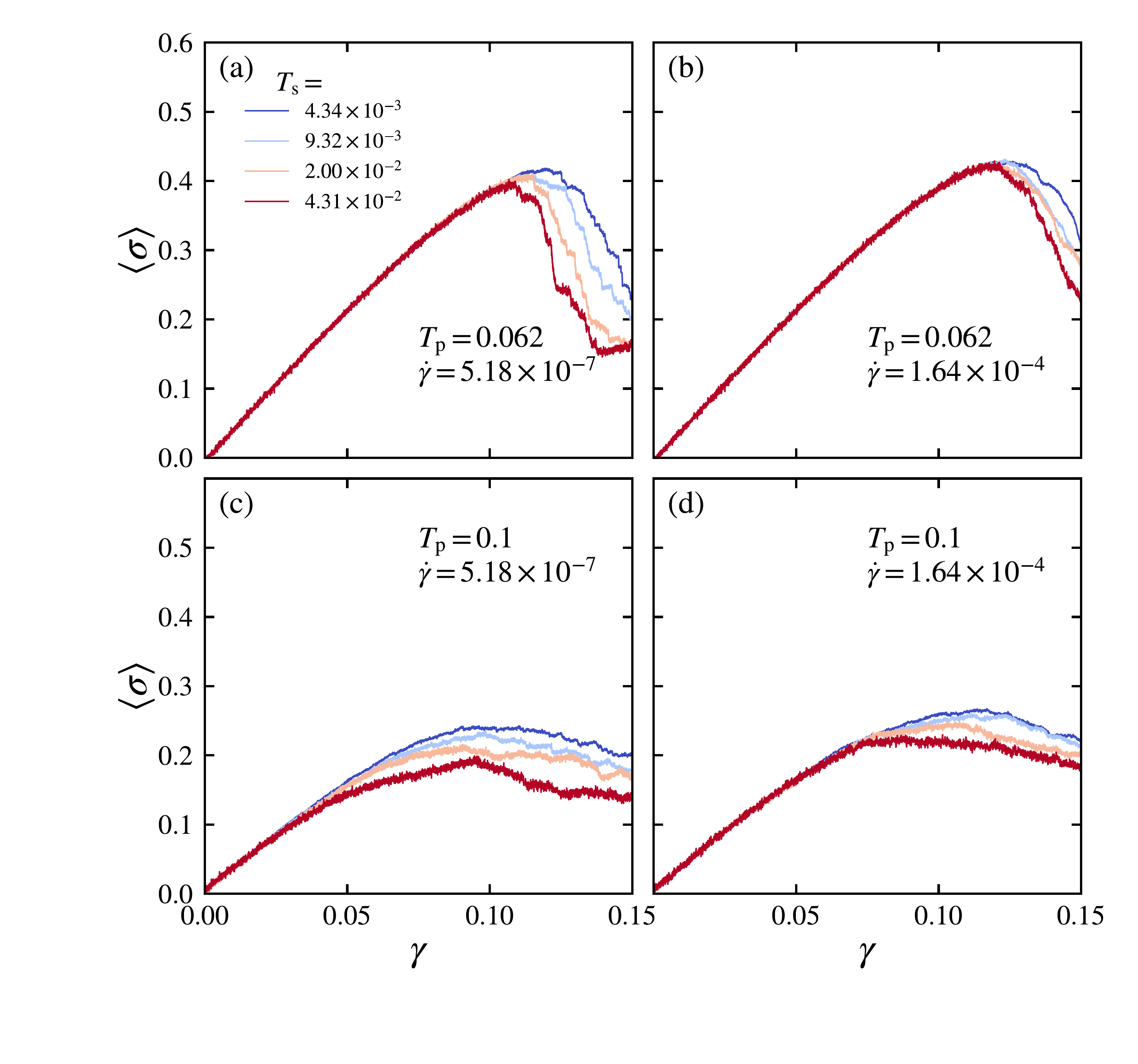}
    \caption{Average stress--strain curves during shear start-up for (a,b) stable glasses, $\Tp=0.062$, and (c,d) more poorly annealed glasses, $\Tp=0.1$. The left and right columns correspond to $\gammadot=5.18\times10^{-7}$ and $\gammadot=1.64\times10^{-4}$, respectively. Colors indicate the temperature $\Ts$, as specified in panel (a).}
    \label{overall_max}
\end{figure}

Figure~\ref{overall_max}(a) shows the average stress--strain curves of the most stable glasses, $\Tp=0.062$, sheared at the low rate $\gammadot=5.18\times10^{-7}$ at several temperatures $\Ts$. All curves follow the same initial elastic response, so the elastic slope does not depend on $\Ts$ in the range studied. Increasing $\Ts$ lowers the start-up yield stress $\sigmamax$ and moves the maximum to smaller strain. After the maximum, the stress drops sharply, the signature of a brittle response. The same trends are observed at the larger rate $\gammadot=1.64\times10^{-4}$ in Fig.~\ref{overall_max}(b), where the curves for different $\Ts$ lie closer together.

Figures~\ref{overall_max}(c) and~\ref{overall_max}(d) show the same protocol for poorly annealed glasses, $\Tp=0.1$. The stress now increases smoothly towards a broad maximum and then relaxes only weakly, without a sharp drop. The response is more ductile. Increasing $\Ts$ mainly lowers the overall stress level. At the larger rate, the stress level is higher and a shallow maximum is more visible, but the relaxation after it remains smooth. The contrast between Figs.~\ref{overall_max}(a,b) and~\ref{overall_max}(c,d) reflects the known effect of glass stability on the nature of yielding, brittle for stable glasses and ductile for poorly annealed ones~\cite{ozawaRandom2018,singhBrittle2020}, which thus remains visible at finite temperature and shear rate. In Secs.~\ref{sec:Rate} and~\ref{sec:Softening}, we analyze how $\sigmamax$ depends on $\gammadot$, $\Ts$ and $\Tp$ for the glasses with $\Tp\leq 0.1$.

\subsection{Shear-rate dependence and zero-rate limit}

\label{sec:Rate}

\begin{figure}[t]
    \centering
    \includegraphics[width=1\textwidth]{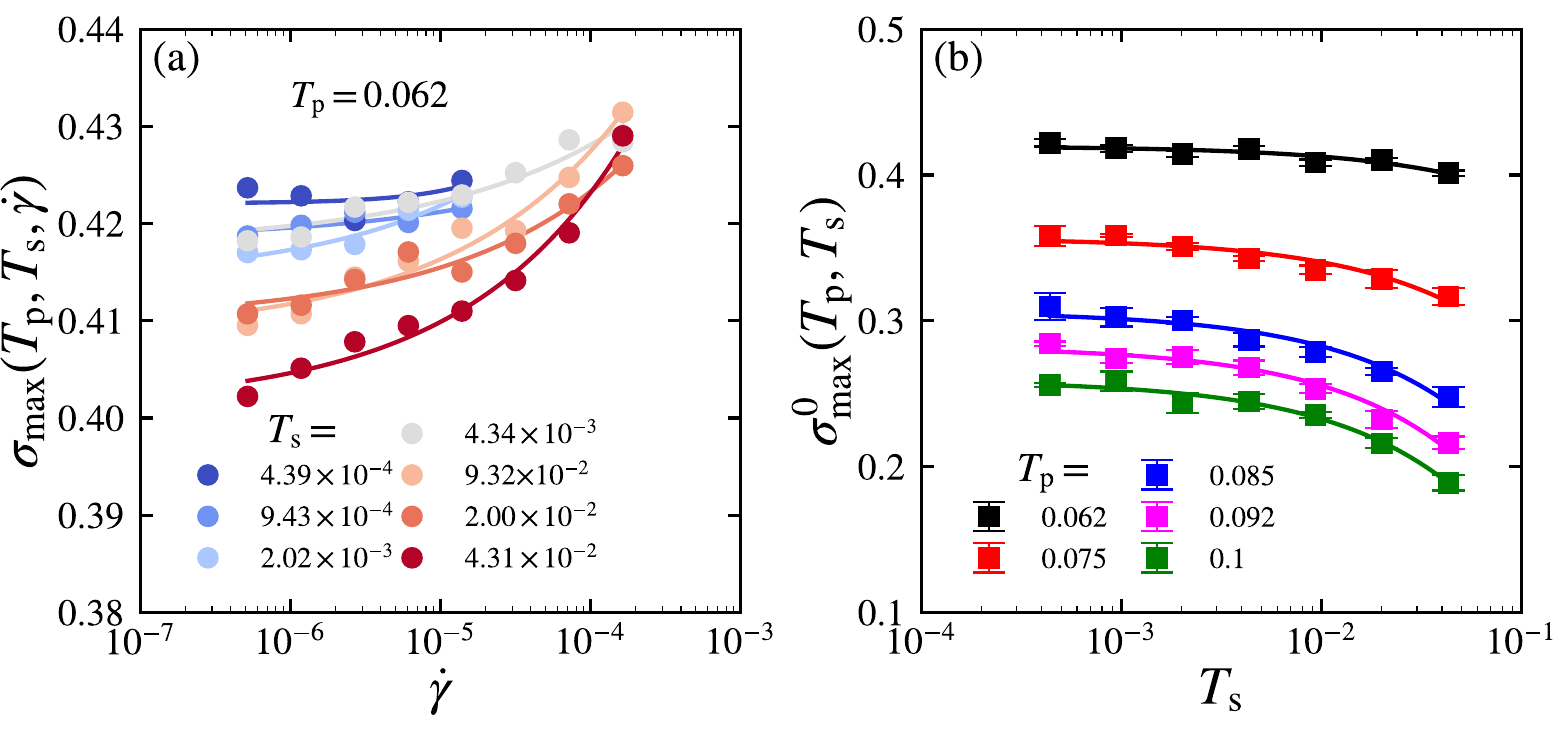}
    \caption{(a) Start-up yield stress $\sigmamax$ as a function of $\gammadot$ for $\Tp=0.062$ and different temperatures $\Ts$. Solid lines are fits to Eq.~\eqref{fit_sigmamax}. (b) Zero-rate start-up yield stress $\sigmamax^{0}$ as a function of $\Ts$ for different parent temperatures $\Tp$. Solid lines are fits to Eq.~\eqref{fit_sigmamax_Ts} with $\mu=2/3$ fixed. Error bars are bootstrap estimates (see Sec.~\ref{sec:Model}).}
    \label{max_fix_Tinit}
\end{figure}

To separate the effects of shear rate and temperature, we first consider the shear-rate dependence of $\sigmamax$ for the most stable glasses, $\Tp=0.062$, shown in Fig.~\ref{max_fix_Tinit}(a). At each $\Ts$, $\sigmamax$ increases with $\gammadot$ and approaches a plateau at low shear rates, whose value decreases with increasing $\Ts$. Experiments and simulations agree that the start-up yield stress depends on $\gammadot$, but there is no generally accepted functional form for this dependence. Depending on the system, it has been described by Herschel--Bulkley~\cite{khabazTransient2021,benziStress2021}, inverse trigonometric~\cite{spaepenMicroscopic1977,chengModelling2021}, power-law~\cite{derecAging2003,divouxStress2011,parkStructural2013,rottlerShear2003} and logarithmic~\cite{rottlerShear2003,varnikStudy2004,shrivastavHeterogeneous2016,golkiaFlow2020} forms. Our curves have the same shape as the steady-state flow curves of Sec.~\ref{sec:SteadyState}, with a plateau at low shear rates followed by an increase at larger $\gammadot$. We therefore use the empirical form
\begin{equation}
    \label{fit_sigmamax}
    \sigmamax(\gammadot) = \sigmamax^{0}+\Kmax\gammadot^{\nmax},
\end{equation}
similar to forms used previously for the start-up yield stress~\cite{rottlerShear2003,benziStress2021}. Here, $\sigmamax^{0}$ is the zero-rate limit of $\sigmamax$, $\Kmax$ is a prefactor and $\nmax$ is an exponent. The fits, shown as solid lines in Fig.~\ref{max_fix_Tinit}(a), describe the data well at high $\Ts$. At lower $\Ts$, the curves are nearly flat over the accessible range of shear rates and the fits are less constrained, similarly to steady state flows. The fits are of similar quality for the other parent temperatures, and the fitted exponents are $\nmax\approx0.4$, with no clear systematic dependence on $\Tp$.

Figure~\ref{max_fix_Tinit}(b) shows the zero-rate start-up yield stress $\sigmamax^{0}$ as a function of $\Ts$ for all parent temperatures $\Tp\leq 0.1$. For every $\Tp$, $\sigmamax^{0}$ decreases with increasing $\Ts$. This shows that thermal fluctuations considerably soften the start-up yield stress even in the limit of vanishing shear rate. At fixed $\Ts$, more stable glasses have a larger $\sigmamax^{0}$, and this ordering is preserved over the whole range of $\Ts$: the glass keeps the memory of its stability at any finite temperature. The amplitude of the softening, however, depends on $\Tp$. It is small for the most stable glasses and grows as $\Tp$ increases. The parent temperature thus controls not only the start-up yield stress, but also its sensitivity to temperature, which we now quantify.

\subsection{Thermal softening}

\label{sec:Softening}

Thermal activation over barriers that vanish at a saddle-node instability predicts that the yield stress decreases as $\Ts^{2/3}$, the Johnson--Samwer law~\cite{johnsonUniversal2005,dasguptaDerivation2013}. Ferrero \textit{et al.}~showed, however, that the thermal exponent depends on the shape of the local energy barriers, the value $2/3$ being recovered for smooth potentials~\cite{ferreroYielding2021}. Since the stability of the glass could modify these barriers, it could in principle also modify the exponent. We therefore describe the temperature dependence of the start-up yield stress, at each shear rate and for its zero-rate limit $\sigmamax^{0}$, by
\begin{equation}
    \label{fit_sigmamax_Ts}
    \sigmamax(\Ts) = \sigmamax^{\ath}-\Amax\Ts^{\mu},
\end{equation}
where $\sigmamax^{\ath}$ is the athermal limit of the start-up yield stress, $\Amax$ is the thermal-softening amplitude and $\mu$ is the thermal exponent. The solid lines in Fig.~\ref{max_fix_Tinit}(b) show that Eq.~\eqref{fit_sigmamax_Ts} describes $\sigmamax^{0}(\Ts)$ well for all $\Tp$.

\begin{figure}[t]
    \centering
    \includegraphics[width=1\textwidth]{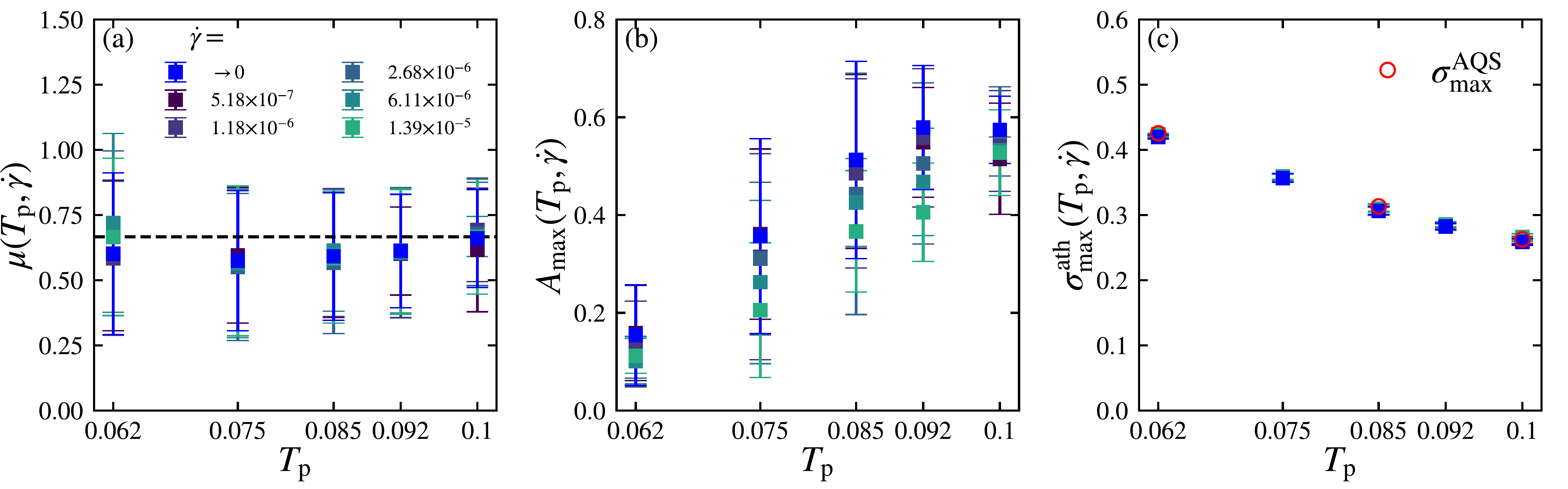}
    \caption{Fit parameters for the temperature dependence of the start-up yield stress to Eq.~\eqref{fit_sigmamax_Ts}, as a function of $\Tp$ for different shear rates $\gammadot$. The label $\gammadot\to0$ refers to the fits of $\sigmamax^{0}(\Ts)$.
    (a) Exponent $\mu$, obtained with all parameters free. The black dashed line indicates the Johnson--Samwer value $\mu=2/3$.
    (b) Thermal-softening amplitude $\Amax$ and (c) athermal start-up yield stress $\sigmamax^{\ath}$, obtained with $\mu=2/3$ fixed. Open circles in (c) indicate the AQS values $\sigmamax^{\AQS}$ from Refs.~\cite{ozawaRandom2018,singhBrittle2020}.
    Error bars are bootstrap estimates (see Sec.~\ref{sec:Model}).}
    \label{sigma_max_shearrate}
\end{figure}

We first leave all three parameters free. The fitted exponent $\mu$, shown in Fig.~\ref{sigma_max_shearrate}(a), is clearly consistent with $\mu= 2/3$ for all parent temperatures and shear rates, with no systematic trend. The uncertainty on each value is not very small, because the accessible range of $\Ts$ does not strongly constrain the exponent. Within these uncertainties, we detect no deviation from $2/3$, such as could arise from logarithmic rate corrections or from barriers of a different shape~\cite{dasguptaDerivation2013,ferreroYielding2021}. We therefore fix $\mu=2/3$ in the following. We have checked that the dependence of $\Amax$ on $\Tp$ discussed below remains qualitatively unchanged if $\mu$ is instead fixed to $0.6$, close to the mean of the fitted values over all $\Tp$ and $\gammadot$.

The resulting thermal-softening amplitude $\Amax$ is shown in Fig.~\ref{sigma_max_shearrate}(b). For all shear rates, $\Amax$ increases markedly with $\Tp$, whereas its variation with $\gammadot$ is weak and comparable to the uncertainties over the range of shear rates studied here, $\gammadot\leq1.39\times10^{-5}$. Less stable glasses are thus much more sensitive to temperature than stable ones. Writing Eq.~\eqref{fit_sigmamax_Ts} as 
\begin{equation}
\sigmamax=\sigmamax^{\ath}\,[1-(\Amax/\sigmamax^{\ath})\,\Ts^{2/3}]
\end{equation}
makes this even clearer. Since $\sigmamax^{\ath}$ decreases with $\Tp$ (see below), the coefficient $\Amax/\sigmamax^{\ath}$, which measures the softening relative to the athermal strength, grows even faster than $\Amax$, from $0.37$ at $\Tp=0.062$ to $2.2$ at $\Tp=0.1$. The effect of stability is therefore not a mere rescaling of the yield stress. Within a thermally activated picture, this trend is consistent with a preparation-dependent distribution of activation barriers. More stable glasses may contain fewer low-barrier regions, or larger typical barriers, and are therefore less likely to yield by thermal activation before the mechanical instability is reached. We emphasize, however, that $\Amax$ is an effective macroscopic amplitude, not a direct measurement of the barrier distribution.

The athermal start-up yield stress $\sigmamax^{\ath}$ is shown in Fig.~\ref{sigma_max_shearrate}(c). It decreases with increasing $\Tp$, reflecting the lower strength of less stable glasses, and shows no systematic dependence on $\gammadot$, within the uncertainties. This weak rate dependence is consistent with the results of Singh \textit{et al.}~for the same model, where the stress--strain response at $\gammadot=10^{-4}$ remains very close to the AQS one, and finite-rate effects only appear for $\gammadot\gtrsim10^{-3}$~\cite{singhBrittle2020}. The open circles in Fig.~\ref{sigma_max_shearrate}(c) show the AQS start-up yield stresses $\sigmamax^{\AQS}$ from Refs.~\cite{ozawaRandom2018,singhBrittle2020}. These values depend only weakly on system size~\cite{ozawaRandom2018,singhBrittle2020}, so they can be compared directly with our results for $N=1500$. The agreement between $\sigmamax^{\ath}$ and $\sigmamax^{\AQS}$ shows that the thermal extrapolation recovers the athermal strength of the glass, as found for the dynamic yield stress in Sec.~\ref{sec:SteadyState}.

\subsection{Comparison with steady flow}

\label{sec:Comp}

\begin{figure}[t]
    \centering
    \includegraphics[width=0.7\textwidth]{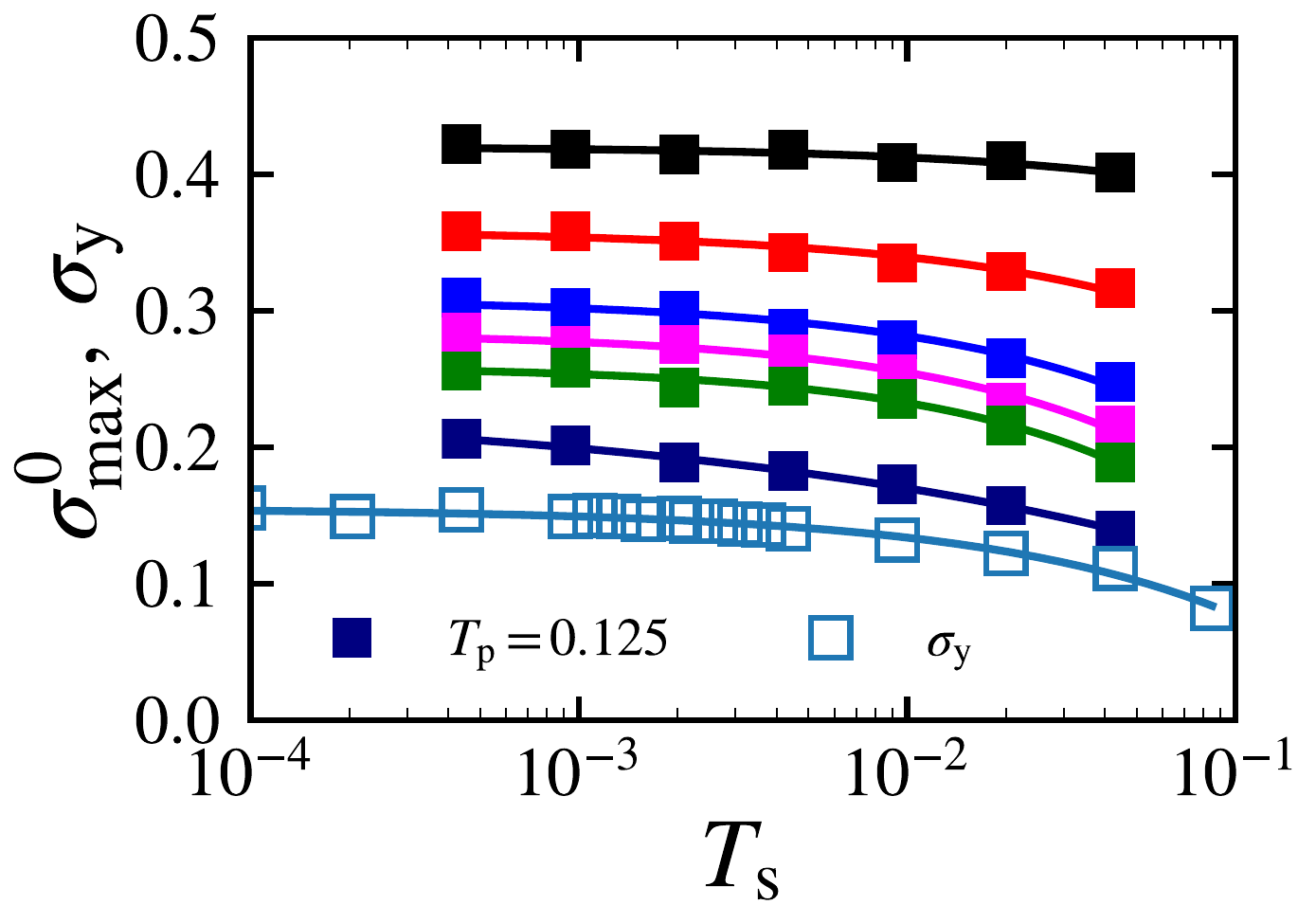}
    \caption{Zero-rate start-up yield stress $\sigmamax^{0}$ for different parent temperatures $\Tp$ (filled squares; colors as in Fig.~\ref{max_fix_Tinit}(b), with $\Tp=0.125$ added in dark blue) and dynamic yield stress $\sigmay$ (open squares), as a function of the temperature $\Ts$. Solid lines are fits to Eq.~\eqref{lemaitre_fit} and to Eq.~\eqref{fit_sigmamax_Ts} with $\mu=2/3$.}
    \label{all_temperature_dependence}
\end{figure}

Figure~\ref{all_temperature_dependence} compares the zero-rate start-up yield stress $\sigmamax^{0}(\Ts)$ of all glasses, including $\Tp=0.125$, for which it is less well defined (Sec.~\ref{sec:Model}), with the dynamic yield stress $\sigmay(\Ts)$ of Sec.~\ref{sec:SteadyState}. All curves decrease with increasing $\Ts$, showing that thermal fluctuations soften both the start-up and the steady-state yield stresses. While $\sigmay$ does not depend, by construction, on the preparation of the glass, $\sigmamax^{0}$ retains the stability hierarchy discussed in Sec.~\ref{sec:Rate}, and lies above $\sigmay$ for all the glasses studied here.

Since both yield stresses decrease as $\Ts^{2/3}$, their sensitivity to temperature can be compared directly through the amplitudes $\Ass$ of Eq.~\eqref{lemaitre_fit} and $\Amax$ of Eq.~\eqref{fit_sigmamax_Ts}, the latter obtained from the fits of $\sigmamax^{0}(\Ts)$. Stable glasses, for which $\Amax<\Ass$, are less sensitive to temperature than the steady flow, whereas poorly annealed glasses, for which $\Amax>\Ass$, are more sensitive. The crossover occurs around $\Tp\approx0.08$. In this respect, the steady flow softens like a glass of intermediate stability. This also explains why the gap between $\sigmamax^{0}$ and $\sigmay$ in Fig.~\ref{all_temperature_dependence} widens with increasing $\Ts$ for the most stable glasses, but narrows for the less stable ones.

Overall, the exponent of the thermal softening is the same for start-up and steady flow, while its amplitude at start-up is controlled by the stability of the glass. As a consequence, the thermal sensitivity measured in steady flow cannot be used to predict that of the start-up yield stress, which can be larger or smaller depending on how the glass was prepared. A description of yielding at finite temperature must therefore account explicitly for the stability of the glass.

\section{Conclusions}

\label{sec:Conclusions}

We have studied how temperature, shear rate and glass stability control the yield stress of a three-dimensional model glass, both in steady flow and at shear start-up, using the swap Monte Carlo algorithm to prepare glasses over a broad range of stability. In steady flow, the dynamic yield stress decreases with temperature as $\Ts^{2/3}$, as predicted by thermal activation near a saddle-node instability, and its athermal extrapolation agrees with the AQS value. At start-up, the zero-rate start-up yield stress also decreases with temperature. Its temperature dependence is compatible with the same $\Ts^{2/3}$ law for all parent temperatures and shear rates, with no systematic dependence of the exponent on stability or shear rate within our uncertainties, and its athermal extrapolation again agrees with the AQS start-up yield stress. By contrast, the amplitude of the thermal softening increases strongly as the glass stability decreases, both in absolute terms and relative to the athermal strength of the glass, while it depends only weakly on shear rate over the range studied. Stable glasses are thus much less sensitive to temperature than poorly annealed ones, and the steady flow softens like a glass of intermediate stability. Glass stability therefore leaves the exponent of the thermal softening unchanged, but strongly controls its amplitude.

These results raise the question of which microscopic property of the initial glass controls the amplitude of the thermal softening. This could be addressed by atomistic simulations of different glass-formers, and by coarse-grained descriptions such as elastoplastic models, where larger system sizes are accessible. More generally, our results call for theoretical descriptions of the thermal softening of the yield stress in glasses of different stability, for instance within soft glassy rheology~\cite{sollichRheology1997,moorcroftAgeDependent2011}, shear transformation zone theory~\cite{langerSheartransformationzone2008}, elastoplastic models~\cite{nicolasDeformation2018,rossiEmergence2022,popovicThermally2021,ferreroYielding2021,korchinskiThermally2024,tahaeiScaling2023}, mode-coupling theory~\cite{braderGlass2009} or random first-order transition theory~\cite{wisitsorasakDynamical2017,agoritsasOutofequilibrium2019}.

Several directions could test the generality of our findings. It would be useful to study the effects of dimensionality, system size and interaction potential, and to determine whether the strong dependence of the softening amplitude on stability persists in other glass-formers~\cite{singhBrittle2020,ozawaRare2022,pollardYielding2022}. Extending the accessible ranges of $\Ts$ and $\gammadot$ would make it possible to test the logarithmic rate corrections expected from thermal activation, and whether the $\Ts^{2/3}$ law remains valid at very low shear rates~\cite{dasguptaDerivation2013,ferreroYielding2021,korchinskiThermally2024}. Finally, experiments on metallic and colloidal glasses could establish the relevance of these trends for real materials~\cite{johnsonUniversal2005,sentjabrskajaTransient2014,divouxStress2011}.

\section*{Acknowledgements}

We thank P. Chaudhuri, K. Martens and A. Lema\^{\i}tre for useful discussions about these resutls. T.W. acknowledges financial support by China Scholarship Council (NO. 202406150002). L.B. acknowledges the support of the French Agence Nationale de la Recherche (ANR), under grants ANR-20-CE30-0031 (project THEMA) and ANR-24-CE30-0442 (project GLASSGO).

\bibliographystyle{unsrt-doilink}
\bibliography{ref}


\end{document}